\documentclass[a4paper,11pt]{article}
\usepackage{graphicx}

\usepackage{times,mathptmx}

\usepackage{latexsym,amssymb,amsthm,amsmath,color
}
\usepackage[hyphens]{url}
\usepackage{bm}
\usepackage{hyperref}
\allowdisplaybreaks
\usepackage{color}
\usepackage{comment}

\title{A conditional match rate anomaly and ranking pressure\\ in residency matching programs\thanks{We are grateful to Fuhito Kojima for his helpful comments and suggestions, as well as to the participants of the 2022 Japanese Economic Association Spring Meeting and the Contract Theory Workshop on Zoom (CTWZ) for their helpful feedback.
This reserch was supported by JSPS KAKENHI Grants (JP17K03776, JP18K01513, JP19K01542, JP22K01402).}}
\author{Munetomo Ando\thanks{Email: ando.munetomo@nihon-u.ac.jp}\\
Nihon University
\and
Minoru Kitahara\thanks{Email: mkitahar@omu.ac.jp}\\Osaka Metropolitan University}
\date{First version: January 26, 2022; Current version: May 13, 2025}

\newtheorem{proposition}{Proposition}

\begin{document}
\maketitle
\begin{abstract}
In the medical residency matching markets of the U.S. and Japan, we observe that an applicant's probability of matching with their first-listed program is disproportionately higher than that of matching with their second-listed program, given that they were rejected by the first. In contrast, the conditional probabilities of matching with lower-ranked programs are markedly lower and remain relatively stable. Furthermore, several experts have noted that participating programs sometimes exert pressure on applicants to manipulate the order of their rank-order lists. 

In this study, we show that this pressure can account for the observed probability pattern, considering the verifiability of being ranked first on the list.
Using empirical data, we identify the prevalence of ranking pressure and quantify its impact on rank-order list changes and welfare under a simplified acceptance and pressure process. Additionally, we explore the implementation of a random permutation of the submitted rank-order list as a measure to counteract list reordering due to pressure. Our analysis shows that the benefits of this intervention outweigh the associated efficiency losses.


\end{abstract}

\section{Introduction}
In many countries, after passing the Medical Licensing Examination and obtaining a medical license, medical graduates are required to attend a residency program. A centralized matching method is used to allocate medical students to residency programs.\footnote{\hyperlink{Roth (2003)}{Roth (2003)} provides a comprehensive overview of the history of residency matching.} In the U.S., the National Resident Matching Program (NRMP) serves as the central clearinghouse, while in Japan, the Japan Residency Matching Program (JRMP) fulfills this role.

Once the necessary procedures are completed, each applicant submits a rank order list to the clearinghouse.\footnote{In 2019, the most recent year prior to the COVID-19 pandemic, the average number of programs on an applicant's list is 9.83 in the U.S. and 3.25 in Japan. The former is a weighted average of 11.22 (matched) and 4.21 (unmatched). See \url{https://www.nrmp.org/wp-content/uploads/2021/08/Impact-of-Length-of-ROL-on-Match-Results-2021.pdf} and \url{https://www.jrmp2.jp/toukei/2019/2019toukei-5.pdf}.} Similarly, each residency program submits a preference list along with a quota, which specifies the maximum number of residents it can accommodate. The clearinghouse then determines the matching results based on an algorithm and communicates the outcomes to the participants.

The clearinghouses in the U.S. and Japan employ a modified version of the Deferred Acceptance Algorithm (DAA) as the matching protocol. The original DAA produces a stable matching, and the students-proposing DAA is strategy-proof for students in the sense that truth-telling is a weakly dominant strategy. 
While the DAA is generally not strategy-proof for the residency programs, it has been shown that truth-telling is approximately optimal in sufficiently large markets (\hyperlink{KP09}{Kojima and Pathak (2009)}).

Nevertheless, narrative and survey-based evidence—widely scrutinized and discussed in the U.S. (see, e.g., \hyperlink{Fisher}{Fisher (2009)}, \hyperlink{Jena}{Jena et al. (2012)}, and \hyperlink{Ree-Jones}{Rees-Jones (2018)})—indicates that students frequently experience pressure from programs to rank them higher than their true preferences during interviews and post‐interview communications.\footnote{For example, in exchange for assuring the hospital that the medical student with whom they are interviewing will be listed within the quota on the list submitted by the hospital, the program requests that the student list it first.}
Consequently, the NRMP explicitly banned such conduct in its Match Participation Agreement,\footnote{https://www.nrmp.org/intro-to-the-match/the-match-agreement/} and established the Match Communication Code of Conduct in 2012 (\hyperlink{Signer and Curtin}{Signer and Curtin 2020}), which mandates that the program directors comply with five clauses, including the following:
\begin{itemize}
    \item \textbf{Respecting an applicant's right to privacy and confidentiality}\\ Program directors and other interviewers may freely express their interest in a candidate, but they shall not ask an applicant to disclose the names, specialties, geographic location, or other identifying information about programs to which the applicant has or may apply.
    \item \textbf{Discouraging unnecessary post-interview communication}\\ Program directors shall not solicit or require post-interview communication from applicants, nor shall program directors engage in post-interview communication that is disingenuous for the purpose of influencing applicants' ranking preferences.\footnote{See the NRMP website (https://www.nrmp.org/communication-code-of-conduct/) for the detail of the code of conduct and see \hyperlink{Sbicca}{Sbicca et al. (2012)} for a case of the code violation.}
\end{itemize}
The JRMP does not explicitly prohibit such behavior; moreover, survey results suggest that this conduct is more widespread in Japan. Specifically, a 2019 survey of participants in the Japanese residency match revealed that 63\% of respondents answered affirmatively to the question, ``Have you ever been asked about your preference ranking during an interview?''\footnote{See \url{https://jrmp2.s3-ap-northeast-1.amazonaws.com/question/2019gakusei.pdf}.}

Even though the current protocol prevents participating programs from directly observing students' submitted lists, a program can still deduce whether a student has succumbed to pressure. Specifically, if a program includes a student within its quota and the student reciprocates by ranking that program first, a match is guaranteed. Consequently, if a program fails to secure a match, it can be deduced that the student did not rank the program first.\footnote{Interestingly, in \hyperlink{Grim}{Grimm et al. (2016)}'s survey, ``52.6\% (141 of 268) of program directors reported that at least once a year 1 or more applicants falsely claim they are ranking their program No. 1.'' This implies that if a program assures a student via post‐interview communication that it will definitely accept the student, the student may feel compelled to rank the program first—a situation distinct from one in which a program pressures a student to rank it higher but not necessarily at the top. In the latter case, even if the student complies, the match may not occur if the student is ultimately matched with a higher-priority program.}\footnote{\hyperlink{Grim}{Grimm et al. (2016)} reported that ``\lbrack i\rbrack n our survey, only 5.2\% (14 of 268) of program directors reported that they always or usually move applicants up their rank order lists after the applicant promises to rank their program No. 1.''}

\begin{figure}[ht]
    \begin{tabular}{c}
          \begin{minipage}[t]{0.5\hsize}
        \centering
        \includegraphics[scale=0.65,width=0.9\linewidth]{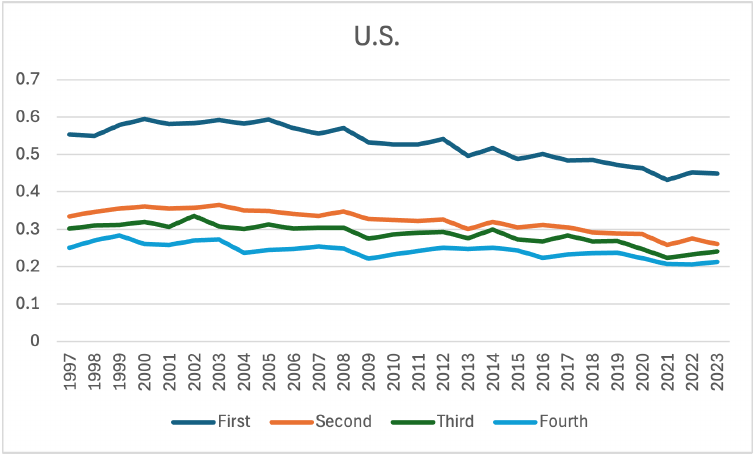}
        \caption{The U.S.}
        \label{us}
      \end{minipage} 
            \begin{minipage}[t]{0.5\hsize}
        \centering
        \includegraphics[scale=0.65,width=0.9\linewidth]{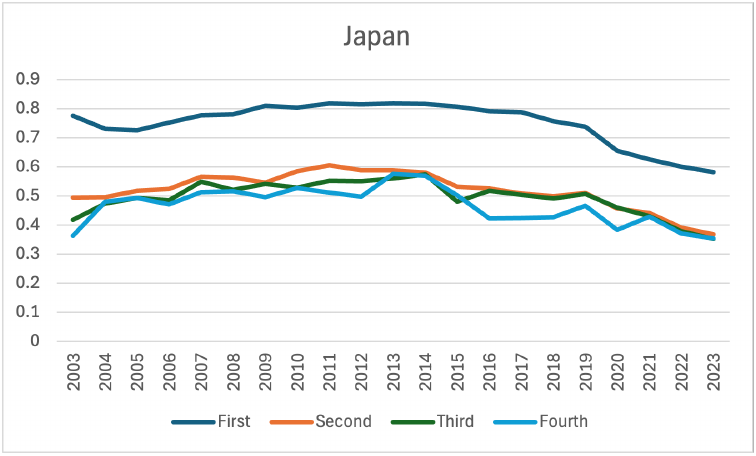}
        \caption{Japan}
        \label{jpn}
      \end{minipage}

    \end{tabular}
  \end{figure}
  
We relate these observations to an anomaly found in both markets.
In each market, data on the number of students assigned to each of the top four programs on their rank order lists is publicly available.\footnote{In both the U.S. and Japan, data on the number of applicants matched to each rank is available. In Japan, additional data on unmatched applicants by list length is also provided, which allows for calculating the acceptance rate at a given list rank conditional on rejection by all higher‐rank programs. 
For example, for Japan, the $k$th rate is calculated as
\begin{equation}
\mbox{the $k$th rate}=\frac{\#\text{matched to the $k$th}}{\#\text{applicants}-\sum_{k'<k}(\#\text{unmatch with the list length of $k'$}+\#\text{matched to the $k'$th})},\nonumber
\end{equation}
and approximately for the U.S. as
\begin{equation}
\mbox{the $k$th rate}\approx\frac{\#\text{matched to the $k$th}}{\#\text{applicants}-\sum_{k'<k}\#\text{matched to the $k'$th}},\nonumber
\end{equation}
as long as the number of unmatch applicants with their list lengths shorter than $k$ is small. Note that for the match probability with the second and subsequent programs on the list, it is necessary to take into account that some applicants may not have enough programs on the list to calculate the probability correctly. Therefore, when we derived the probabilities for Japan, we properly dealt with this issue in our calculations. However, we did not take into account it when calculating the probabilities for the U.S. due to data limitations.} Consequently, by analyzing a matching process where a student is assigned to a program only if all higher-ranked programs reject the student and the program accepts the student, we can compute the match rate for each rank, conditional on prior rejections. As shown in Figures 1 and 2, the conditional match rate for the second rank exhibits a sharp decline relative to the first rank, while the rates for subsequent ranks remain relatively stable.\footnote{Figures \ref{us} and \ref{jpn} show a conditional match rate anomaly. These figures tell us that an applicant's match probability with the program listed first on the list is consistently high. In the U.S., an applicant has approximately a 50\% probability of being matched with the first program on the list. In Japan, the probability is approximately 75\%.}

The observed change in conditional match rates between the first and subsequent ranks can be explained as follows. 
Effective pressure that secures a match arises only when a program that exerts pressure is ranked first by a student and is willing to accept the student, as discussed earlier. Consequently, the match rate for the first-listed program is elevated. In contrast, a rejection by the first-listed program implies that no effective pressure was applied, which uniformly reduces the conditional match rates for the second and lower ranks.

Since the conditional match rates for the second through fourth ranks are nearly constant, it is reasonable to assume that, in each market, both the probability of a program accepting a student and the probability of a program exerting pressure on an acceptable student are independent of the program's position on the student's true preference list. As an initial step in quantification, we calibrate these values to estimate the extent to which pressure influences changes in the submitted list, thereby complementing existing narrative and survey evidence.

This raises the following question: To what extent does pressure from programs on applicants to modify their lists reduce matching efficiency?

The match outcome can be altered if an applicant, due to external pressure, submits a list that deviates from their true preference order. For instance, if an applicant had submitted a list reflecting their true preferences, they might have been matched with their first-choice program; however, they were instead matched with their third-choice program. This difference of two ranks is defined as the number of rank losses caused by the forced list change and serves as a measure of welfare loss.

A high likelihood of forced list modification does not necessarily result in substantial welfare loss. Specifically, if all programs that find an applicant acceptable exert pressure, then—as long as the applicant ranks the most preferred among them first—the match outcome remains unchanged (see Remark 1). Thus, it is crucial to empirically assess not only the probability of applicants experiencing pressure but also the extent of the resulting welfare loss. Data from the U.S. and Japan indicate that both the incidence of pressure and the resulting welfare loss are significant under a simplified reduced-form acceptance and forcing process (Section 3).\footnote{For analytical tractability, we consider the following reduced-form process. Applicants are assumed to be ex-ante homogeneous. The programs simultaneously decide whether the applicants are acceptable with an exogenous acceptance probability, $a$. We also assume that each acceptant program exerts pressure on the applicant with an exogenous probability, $e$. The applicant who receives effective pressure from at least one program will choose the best program and place it first on the list.}

How can these losses be mitigated? Applicants alter their rank order lists under external pressure due to the risk that failing to honor a promise—and consequently not securing a match with the program—may lead to reputational damage. Consequently, reducing the strong link between promise violations and match failures in the current procedure may be beneficial.

We consider implementing a random, unobservable exchange between the first and second positions on the applicants' submitted rank order lists with positive probability (Section 2.1.5). By introducing such an intervention, it becomes infeasible to ascertain whether an applicant has breached their commitment, as match outcomes are no longer deterministic.

Furthermore, even if an applicant adheres to their commitment, there exists a probability—above a given threshold—that they will be erroneously perceived as having violated it. As long as this threshold remains below 0.5, an adequately high exchange probability can guarantee that this probability approaches the threshold (see Remark 2). Consequently, by allowing applicants to submit their true preference rankings, this intervention fully mitigates any welfare loss induced by external pressure.

However, the random exchange mechanism introduces its own welfare loss; thus, the net impact of this intervention must be empirically assessed. By re-examining our simplified model, we demonstrate that the benefits of the proposed intervention surpass its associated costs, utilizing data from both the U.S. and Japan. If the threshold is set at 15\%, the expected reduction in rank loss is approximately 42\% in Japan (from 0.2053 to 0.0867) and around 6\% in the U.S. (from 0.7740 to 0.0485) (see Section 3).

The rest of the paper is structured as follows. In Section 1.1, we review the related literature. Section 2 presents the model and the main results. Section 3 provides the empirical evaluations. Section 4 examines various issues, including the potential for strategic manipulation, and Section 5 provides concluding remarks. 

\subsection{Related literature}
This paper studies a simplified model of residency matching. In the doctor-proposing DAA, the matching mechanism is renowned for being strategy-proof for the applicants. Nevertheless, instances of applicants deliberately misreporting their preferences have been widely documented. \hyperlink{Hassidim}{Hassidim et al. (2017)}  discuss various justifications for applicants' preference misrepresentation under the truthful mechanism, including failure to identify the dominant strategy and mistrust of the market maker.
Furthermore, \hyperlink{Ashlagi}{Ashlagi and Gonczarowski (2018)} note that the DAA is not an ``obviously strategy-proof mechanism'' as defined by \hyperlink{Li}{Li (2017)}.

We examine the potential influence exerted by residency programs on applicants during the matching process. \hyperlink{Roth and Xing}{Roth and Xing (1994)} noted this possibility as ``\lbrack s\rbrack ome of the most widely reported practices, such as the efforts of employers to extract from students pledges to rank them first, [\ldots].''(p.1022). \hyperlink{Sonmez}{S\"{o}nmez (1999)} focuses on prior agreements that benefit both sides rather than directly addressing pressure.

\hyperlink{Rees-Jones}{Rees-Jones (2018)} reported that, based on data from 579 respondents across 23 U.S. medical schools in 2012, 16\% of respondents submitted rank order lists that deviated from their true preferences in a manner that was not merely attributable to error.

Rees-Jones (2018) discussed the presence of pressure in footnote 8. However, he reports that only 16 (3\%) respondents in his survey have ever been offered a deal whereby the program would also place the student at the upper portion of its own list in exchange for the student placing the program at the higher position, and only three students have actually changed their listings.

Thus, Rees-Jones (2018) concluded that a small but significant amount of pressure-based trading exists. However, because surveys alone may be insufficient to capture the sensitive nature of pressure, our study—which demonstrates the presence of pressure and quantifies its frequency using published data—complements his findings.

Our paper explains the observed matching probabilities for applicants' first- through fourth-listed programs (see Figures 1 and 2). Recently, \hyperlink{Echenique}{Echenique et al. (2022)} examined the matching process in detail and investigated why the matching probability for first-listed programs is very high; however, they did not address the discrepancy between the match probability of the first-listed program and those of lower-ranked programs.

We consider implementing a random permutation of the submitted list to prevent pressure-induced list reordering and the resulting welfare loss. This intervention aligns with the concept of differential privacy, which has recently attracted considerable attention in computer science. Differential privacy originated with \hyperlink{Dwork}{Dwork (2006)}, and a simple implementation known as the Post Randomisation Method (PRAM) is studied by \hyperlink{Kooiman}{Kooiman et al. (1997)}.

Recent economic research has explored the application of differential privacy in mechanism design. For example, \hyperlink{McSherry}{McSherry and Talwar (2007)} show that differential privacy can be a useful tool for achieving incentive compatibility and promoting truthful behavior among agents. In this context, our study examines how matching theory is modified when incorporating methods inspired by differential privacy and quantifies the trade-offs associated with privacy-preserving interventions.

\hyperlink{Pourbabaee}{Pourbabaee and Echenique (2023)} examine a model of public-goods provision that incorporates a randomized "flopped" preference reporting mechanism to address privacy concerns. Their work and ours are related in that both address privacy issues using the concept of random replacement.

\section{Model}
We consider a representative doctor who may be matched with at most one residency program in $I$.
Here $I=\{1,\dots,L\}$ with $L\geq2$ is the set of programs that the doctor has applied to and found to be acceptable, where program $k$ denotes the doctor's $k$th most preferred program.
Thus, we implicitly assume that there is no tie in ranking.

\subsection{Process}
\subsubsection{Matching}
Given the doctor's list of programs $\bm{l}=(l_1,\dots,l_L)\in\{(l'_1,\dots,l'_L)\in I^L|l_{k}\neq l_{k'}\text{ for }k'\neq k\}$,
where program $l_k$ is the program listed $k$th in the list, and the set of programs $A\subset I$ which would accept the doctor, the doctor is matched with program $\mu(\bm{l},A)=l_k$, where $k$ is the smallest index such that $l_k\in A$. If no such $k$ exists, the doctor remains unmatched.

Note that the elements of set $A$ generally depend on other doctors' lists. However, we ignore the possibility in our model.
In this sense, our model is just a simplified reduced-form one.

We also note that the doctor's list $\bm{l}$ is not necessarily equal to their preferences, $I=\{1,\dots,L\}$. This is because, as we will see below, the doctor may alter their submitted list if they feel pressured by the programs. Moreover, as we will study later, the list $\bm{l}$ may further be different from the submitted list, because interventions implemented by the clearinghouse may result in a random permutation of the submitted list (Section 2.1.5). 

In summary, there are three kinds of orders for a particular applicant: the order by true preference, the order in the list submitted to the clearinghouse, and the order in the list used in the matching algorithm after random permutation.

\subsubsection{Ranking pressure}
Given a set of programs exerting effective pressure on the doctor to be listed first, denoted $F\subset\{1,\dots,L\}$, the doctor's submitted list under the pressure structure $F$, denoted $\lambda(F)=(\lambda_1(F),\dots,\lambda_L(F))$, is constructed as follows:
\begin{itemize}
\item If $F=\emptyset$, $\lambda(F)=(1,\dots,L)$, i.e., the doctor submits their true preference list.
\item If $F\neq\emptyset$, $\lambda_1(F)=\min\{i\in F\}$,  and the remaining programs follow in the original preference order, excluding $\lambda_{1}(F)$.
\end{itemize}

We assume the doctor always chooses the most preferred among pressuring programs, representing a best-case scenario for welfare loss. In reality, a doctor might choose a less-preferred program under pressure, for instance, if a program reached out earlier than others. Therefore, our later welfare loss estimates should be interpreted as lower bounds and may underestimate the potential benefits of eliminating such pressures.

\subsubsection{Pressure and acceptance}
Let $e\in(0,1)$ denote the probability with which each program that would accept the doctor independently exerts effective pressure on them. Also, the probability that a program that does not accept the doctor will exert pressure is zero.
In summary, we denote the pressure structure by $\tilde{F}^e(A)$, which is a random subset of $A$ where each program $i\in A$ is included independently with probability $e$.

Note that, by letting the probability be zero if the program $i$ is not an element of the set $A$, we ignore the possibility that a program, uncertain about whether it will ultimately accept the doctor, may still succeed in putting effective pressure on them.\footnote{Including the possibility that programs not in $A$ (i.e., those that would not accept the doctor) exert pressure may be important. For example, as noted by \hyperlink{Doolittle}{Doolittle (2017)}, applicants may misinterpret benign signals from programs as pressure. This possibility can be incorporated by allowing $i\in\tilde{F}^e(A)$ with some small probability $\underline{e}\in(0,e]$ even when $i\notin A$. The difference between models with $\underline{e}=0$ and $\underline{e}>0$ helps capture the effect of misunderstanding. In particular, the conditional match probability after rejection by the first choice may differ, as discussed below.\label{mu}}
By the identicalness and independence, we also ignore that the decisions of the doctor and a program regarding the pressure in general depend on how the doctor evaluates the program and whether other programs also put pressure.

\subsubsection{Acceptance process}
We assume that each residency program accepts the doctor independently with a fixed probability $a\in(0,1)$.
We denote as $\tilde{A}^a$ the set of programs that accept the doctor, that is, $i\in\tilde{A}^a$ with probability $a\in(0,1)$, independently.

Since we assume homogeneity, we ignore the possibility of a correlation between the doctor's preferences and each program's preferences over doctors, such as one arising from the doctor's fit with a particular program.\footnote{In fact, by assuming an extreme correlation distribution, we can reproduce the conditional match rate pattern without resorting to the pressure process, which may actually be plausible to explain just one observation.
However, this pattern is observed in both the U.S. and Japanese markets, as well as across different time periods, with such varying magnitudes. Therefore, it is unrealistic to assume that the parameters of the extreme distribution happen to take appropriate values in each case merely by chance.}
By the assumption of independence, we also ignore the possibility of some doctor generally being prioritized over another doctor, such as that caused by the former having higher ability than the latter.

Then, by letting $\tilde\lambda^{a,e}=(\tilde\lambda^{a,e}_{1},\cdots,\tilde\lambda^{a,e}_{L})\equiv\lambda(\tilde{F}^e(\tilde{A}^a))$ for notational simplicity, the event that a program $i$ is the $k$th in the submitted list is represented by $i=\tilde\lambda^{a,e}_k$.

\subsubsection{Random exchange}
We also consider randomly permuting the doctor's submitted list $\bm{l}=(l_1,l_2,l_3,\dots,l_L)$ before running the matching algorithm.
In particular, for $\epsilon\in[0,1/2]$,
let $\tilde\pi^\epsilon(\bm{l})=(l_2,l_1,l_3,\dots,l_L)$ with probability $\epsilon$ and $\bm{l}$ with probability $1-\epsilon$.\footnote{This paper examines a simple method of changing the doctor's submitted list. Naturally, there can be other effective ways to change the list. It is also important to consider the measures that the programs can take against such intervention methods. For example, a program could not only ask a doctor to write it first on the list, but also request that the second and lower places on the list be left blank.}
That is, with probability $\epsilon>0$, we consider swapping the first and the second choices in the submitted list.\footnote{When the exchange probability $\epsilon$ exceeds 1/2, a doctor could preemptively switch their first and second choices to counteract the expected permutation. Hence, we restrict our attention to cases where $\epsilon \leq 1/2$.} and then, given a submitted list $\bm{l}$ and an acceptance structure $A$, the resulting matching becomes $\tilde\mu^\epsilon(\bm{l},A)\equiv\mu(\tilde\pi^\epsilon(\bm{l}),A)$.
Note that if $\epsilon=0$, then they are the original list and procedure, $\tilde\pi^0(\bm{l})=\bm{l}$ and hence $\tilde\mu^0(\bm{l},A)=\mu(\bm{l},A)$.

\subsection{Key statistics and remarks}
\subsubsection{Conditional match rates}
Let $P_k^{a,e}$ denote the $k$th {\it conditional match rate}, i.e., the probability that a doctor is matched with the $k$th program, conditional on being rejected by all programs ranked higher.
That is,
$P_1^{a,e}\equiv\Pr\left(\tilde\lambda^{a,e}_1\in\tilde{A}^a\right)$,
and for all $k\in\{2,\dots,L\}$,
\begin{equation}
P_k^{a,e}\equiv\Pr\left(\tilde\lambda_k^{a,e}\in\tilde{A}^a\left|\tilde\lambda_{k'}^{a,e}\notin\tilde{A}^a\text{ for all }k'<k\right.\right).\nonumber
\end{equation}

Then, we show that the conditional match rate is highest for the top-ranked program and becomes constant from the second rank onward, as observed in the U.S. and the Japanese markets.
\newline

\noindent{\bf Proposition.} {\it %$P^{a,e}_1=1-(1-ae)^L\frac{1-a}{1-ae}$, and for all $k\in\{2,\dots,L\}$, $P^{a,e}_k=\frac{a-ae}{1-ae}<P^{a,e}_1$
$P^{a,e}_1>P^{a,e}_2=\cdots=P^{a,e}_L$.}
\newline

\noindent{\bf Proof.}
We show that $P^{a,e}_1=1-(1-ae)^L\frac{1-a}{1-ae}$, which is larger than $1-(1-a\times0)\frac{1-a}{1-a\times1}=a$, and for all $k\in\{2,\dots,L\}$, $P^{a,e}_k=\frac{a-ae}{1-ae}$, which is smaller than $\frac{a-a\times0}{1-a\times0}=a$.
For notational simplicity, let $\tilde{F}^{a,e}\equiv\tilde{F}^e\left(\tilde{A}^a\right)$.

For the former, note that $\tilde\lambda_1^{a,e}\in\tilde{A}^a$ if $\tilde{F}^{a,e}\neq\emptyset$,
since $\Pr\left(i\in\tilde{F}^e(A)|i\notin A\right)=0$, it follows that $i\in\tilde{F}^e(A)$ implies $i\in A$,
and that $\tilde\lambda^{a,e}=(1,\dots,L)$ if $\tilde{F}^{a,e}=\emptyset$.
Thus, $P^{a,e}_1=1-\Pr\left(\tilde\lambda^{a,e}_1\notin\tilde{A}^a\right)$ is equal to
\begin{equation}
1-\Pr(\tilde{F}^{a,e}=\emptyset)\Pr(1\notin\tilde{A}^a|\tilde{F}^{a,e}=\emptyset)=1-\Pr(\tilde{F}^{a,e}=\emptyset)\Pr(1\notin\tilde{A}^a|1\notin\tilde{F}^{a,e}),\nonumber
\end{equation}
where we use the conditional independence for the last equality.

For the latter,
by $\Pr\left(\tilde\lambda^{a,e}_1\notin\tilde{A}^a\left|\tilde{F}^{a,e}\neq\emptyset\right.\right)=0$,
$P^{a,e}_k=\frac{\Pr\left(\tilde\lambda^{a,e}_{k'}\notin\tilde{A}^a\text{ for all }k'<k\text{ but }\tilde\lambda^{a,e}_k\in\tilde{A}^a\right)}{\Pr\left(\tilde\lambda^{a,e}_{k'}\notin\tilde{A}^a\text{ for all }k'<k\right)}$ is equal to
\begin{equation}
\frac{\Pr\left(\tilde\lambda^{a,e}_{k'}\notin\tilde{A}^a\text{ for all }k'<k\text{ but }\tilde\lambda^{a,e}_k\in\tilde{A}^a\left|\tilde{F}^{a,e}=\emptyset\right.\right)}{\Pr\left(\tilde\lambda^{a,e}_{k'}\notin\tilde{A}^a\text{ for all }k'<k\left|\tilde{F}^{a,e}=\emptyset\right.\right)},\nonumber
\end{equation}
which is equal to, since $\tilde\lambda^{a,e}=(1,\dots,L)$ if $\tilde{F}^{a,e}=\emptyset$,
\begin{align}
&\frac{\Pr\left(k'\notin\tilde{A}^a\text{ for all }k'<k\text{ but }k\in\tilde{A}^a\left|\tilde{F}^{a,e}=\emptyset\right.\right)}{\Pr\left(k'\notin\tilde{A}^a\text{ for all }k'<k\left|\tilde{F}^{a,e}=\emptyset\right.\right)}\nonumber\\
&=\frac{\Pr\left(k'\notin\tilde{A}^a\text{ for all }k'<k\text{ but }k\in\tilde{A}^a\right)\Pr\left(\tilde{F}^{a,e}=\emptyset\left|k'\notin\tilde{A}^a\text{ for all }k'<k\text{ but }k\in\tilde{A}^a\right.\right)}{\Pr\left(k'\notin\tilde{A}^a\text{ for all }k'<k\right)\Pr\left(\tilde{F}^{a,e}=\emptyset\left|k'\notin\tilde{A}^a\text{ for all }k'<k\right.\right)}\nonumber\\
&=\frac{\Pr\left(k\in\tilde{A}^a\right)\Pr\left(k\notin\tilde{F}^{a,e}\left|k\in\tilde{A}^a\right.\right)}{\Pr\left(k\notin\tilde{F}^{a,e}\right)},\nonumber
\end{align}
where we use the conditional independence for the last equality.
\hfill$\Box$
\newline

That is, since the programs that exert pressure are exactly those that would accept the doctor, and the doctor lists one of them first whenever such a program exists, a rejection from the top-ranked program implies that none of the remaining programs both exert pressure and would accept the doctor. Because the preference list remains unchanged in this case,\footnote{The critical assumption is that programs who would not accept the doctor cannot put the effective pressure.
See also footnote \ref{mu}.} the conditional match rates from the second rank onward are uniformly reduced, from $\Pr(i\in A)$ to $\Pr(i\in A|i\notin \tilde{F}^e(A))$.
Moreover, the cases where the most preferred program would not accept the doctor but there is one which puts the pressure are added to the first match cases, increasing the probability from $\Pr(i\in A)$
to $1-\Pr\left(\tilde{F}^{a,e}=\emptyset\right)\Pr\left(1\notin\tilde{A}^a|1\notin\tilde{F}^{a,e}\right)$.

Note that, given $L$, we can recover $a$ and $e$ from the values of $P^{a,e}_1$ and $P^{a,e}_2$.
In particular,
since $1-P^{a,e}_2=1-\frac{a-ae}{1-ae}=\frac{1-a}{1-ae}$, $1-ae=\left(\frac{1-P^{a,e}_1}{\frac{1-a}{1-ae}}\right)^{1/L}=\left(\frac{1-P^{a,e}_1}{1-P^{a,e}_2}\right)^{1/L}$,
and hence
\begin{equation}
a=1-(1-ae)\left(1-P^{a,e}_2\right)=1-\left(\frac{1-P^{a,e}_1}{1-P^{a,e}_2}\right)^{1/L}\left(1-P^{a,e}_2\right),\label{a}
\end{equation}
by which
\begin{equation}
e=\frac{1-(1-ae)}a=\frac{1-\left(\frac{1-P^{a,e}_1}{1-P^{a,e}_2}\right)^{1/L}}{1-\left(\frac{1-P^{a,e}_1}{1-P^{a,e}_2}\right)^{1/L}(1-P^{a,e}_2)}.\label{e}
\end{equation}
We will use them to empirically evaluate the U.S. and the Japanese markets in Section 3.

\subsubsection{Rank loss}
This subsection examines the rank loss caused by external pressure that leads the doctor to change the submitted preference list. We define rank loss as the difference in rank between the program assigned under the submitted (possibly manipulated) list and the program that would have been assigned under the doctor’s truthful preference list.\footnote{\hyperlink{Featherstone}{Featherstone (2020)} is a study that focuses on rank distribution when assessing the quality of matching results.}

Let $RL^{a,e}$ denote the average rank loss resulting from the forced list change. That is,
\begin{equation}
RL^{a,e}\equiv E\left[\mu\left(\tilde\lambda^{a,e},\tilde{A}^a\right)-\mu\left((1,\dots,L),\tilde{A}^a\right)\right].\nonumber
\end{equation}
We define the difference to be zero when $\tilde{A}^a=\emptyset$.\footnote{Note that $\mu\left(\lambda(\tilde{F}^e(A),A\right)=\emptyset$ if and only if $\mu\left((1,\dots,L),A\right)=\emptyset$, which is equivalent to $A=\emptyset$.}

We use this as a measure of the welfare loss.
Note that this measure overlooks many welfare-related considerations.
For example, the impact of a one-rank loss may vary depending on the level of the rank, which could be addressed by introducing a non-linear utility function $u(i)$ over ranks.
Moreover, the existence of pressure itself may be stressful (see, e.g., \hyperlink{Doolittle}{Doolittle (2017)}), even if the resulting match is unchanged, for which some evaluation of $\tilde{F}^{a,e}$ itself, such as $-c|\tilde{F}^{a,e}|$ with $c>0$ being the constant stress cost, is necessary.

It is noteworthy that although the initial drop in the conditional match rate increases with the pressure probability $e$, the average rank loss, which initially rises from zero, eventually vanishes as pressure becomes nearly certain.
\newline

\noindent{\bf Remark 1.} {\it While $P^{a,e}_1-P^{a,e}_2$ is increasing in $e$, $RL^{a,e}>\lim_{e\downarrow0}RL^{a,e}=\lim_{e\uparrow1}RL^{a,e}=0$.}
\newline

\noindent{\bf Proof.}
The former part simply follows from that $P^{a,e}_1=1-(1-ae)^L\frac{1-a}{1-ae}$ is increasing, and $P^{a,e}_2=\frac{a-ae}{1-ae}$ is decreasing in $e$.

For the latter, note that $RL^{a,e}$ is equal to
\begin{equation}
\sum_{i=1}^{L-1}\Pr\left(\min_{i'\in\tilde{A}^a}i'=i\right)\Pr\left(\min_{i'\in\tilde{F}^{a,e}}i'>i\left|\min_{i'\in\tilde{A}^a}i'=i\right.\right)E\left[\min_{i'\in\tilde{F}^{a,e}}i'-i\left|\min_{i'\in\tilde{F}^{a,e}}i'>\min_{i'\in\tilde{A}^a}i'=i\right.\right],\nonumber
\end{equation}
the second term of which is equal to, by the conditional independence,
\begin{equation}
\Pr\left(i\notin\tilde{F}^{a,e}\left|i\in\tilde{A}^a\right.\right)\left(1-\prod_{i'>i}\Pr\left(i'\notin\tilde{F}^{a,e}\right)\right)=(1-e)(1-(1-ae)^{L-i}),\nonumber
\end{equation}
which converges, by $i<L$, to 0 as $e$ approaches either 0 or 1.
\hfill$\Box$
\newline

In other words, when the probability of pressure conditional on acceptance is close to one, the absence of pressure from a program almost surely implies that the program would not have accepted the doctor anyway. Therefore, lowering the rank of such a program in the list has negligible consequences.
Moreover, as long as the doctor chooses the best among the pressured programs as the first listed, the match outcome does not change.

This implies that a substantial drop in the conditional match rate, as observed in the U.S. and Japanese markets, may coexist with minimal rank loss.
Hence, it remains an empirical question whether the degree of pressure sufficient to cause such a large observed gap in match rates also results in a meaningful welfare loss -- an issue we address in the empirical section.

\subsubsection{Type I errors}
We examine whether a program that pressures a doctor to rank it first can detect whether the doctor actually complied.

Let $Q^{a,e,\epsilon}_{i\to k}$ denote the probability that program $i$ is not matched with the doctor, conditional on (i) the program being willing to accept the doctor, and (ii) the doctor listing the program in the $k$th position due to pressure, for which we further specify that the orders among others would remain the same in the changed list.
That is, let $\tilde\lambda^{a,e}_{i\to k}$ be such that $\tilde\lambda^{a,e}_{i\to k,k}=i$ and for all $k',k''\in\left\{k'''\in\{1,\dots,L\}\left|\lambda^{a,e}_{k'''}\neq i\right.\right\}$, $\tilde\lambda^{a,e}_{i\to k,k'}>\tilde\lambda^{a,e}_{i\to k,k''}$ if and only if $\tilde\lambda^{a,e}_{k'}>\tilde\lambda^{a,e}_{k''}$.
Then,
\begin{equation}
Q^{a,e,\epsilon}_{i\to k}=\Pr\left(\tilde\mu^\epsilon\left(\tilde\lambda^{a,e}_{i\to k},\tilde{A}^a\right)\neq i\left|i\in\tilde{F}^{a,e}\cap\tilde{A}^a\right.\right).\nonumber
\end{equation}
Note that $Q^{a,e,0}_{i\to k}$ is the probability under the original procedure, with no addition of permutation.

We consider that (i) with a significance level $\alpha>0$, conditional on that the program is not matched with the doctor, the hypothesis that the doctor follows the pressure to list $k$th a program that would accept the doctor is rejected if the type I error rate is below the significance level, $Q^{a,e,\epsilon}_{i\to k}\leq\alpha$, and (ii) this rejection possibility makes the doctor follow the pressure.
Then, under the original procedure ($\varepsilon = 0$), listing a program first has a distinctive property: the hypothesis that the doctor followed the pressure can be rejected at any arbitrarily small significance level—that is, the type I error is zero. In contrast, for all other ranking positions, the minimum significance level required for rejection is strictly positive, i.e., bounded away from zero.
\newline

\noindent{\bf Remark 2.} $Q^{a,e,0}_{i\to 1}=0\leq\alpha$ for any $\alpha$. Moreover, for all $k\geq2$, $Q^{a,e,0}_{i\to k}\geq %Q^{a,e,0}_{i\to 2}=
\underline{Q}^{a,e}\equiv1-(1-a)(1-ae)^{L-2}$.
\newline

\noindent{\bf Proof.}
%The former follows from the definition.
By $\tilde\mu^0=\mu$, $\tilde\lambda^{a,e}_{i\to 1,1}=i$ and $i\in\tilde{A}^a$ imply $\tilde\mu^0(\tilde\lambda^{a,e}_{i\to 1},\tilde{A}^a)=i$,
which proves the former.
Moreover,
for the latter, for $k\geq2$, $Q^{a,e,0}_{i\to k}$ is equal to
\begin{align}
&1-\Pr\left(\tilde{A}^a\cap\left\{1,\dots,k-\mathbf1_{i\geq k}\right\}\setminus\{i\}=\emptyset\right)\Pr\left(\tilde{F}^{a,e}\cap\left\{k+\mathbf1_{i<k},\dots,L\right\}\setminus\{i\}=\emptyset\left|\tilde{A}^a\cap\left\{1,\dots,k-\mathbf1_{i\geq k}\right\}\setminus\{i\}=\emptyset\right.\right)\nonumber\\
&=1-(1-a)^{k-1}(1-ae)^{L-k},\label{qae0}
\end{align}
where we use the conditional independence for the last equality.\footnote{Here $\mathbf1_{x}$ is the indicator function that is eqaul to 1 if $x$ holds; 0 otherwise.}
\hfill$\Box$
\newline

That is, since the program would accept the doctor, as long as the doctor truly listed the program first, the program would be certainly matched with the doctor.
However, for other ranks, even if the doctor follows the listing promise, the unmatch may occur because the doctor may match another program listed higher,
the probability of which is enhanced by the existence of first listing pressure which makes programs listed higher more probable to accept the doctor.
Of course, it may be negligible if the lower bound itself is positive but so small as standard significance levels, such as 10\%,
which will be checked after the estimation of $a$ and $e$.

\subsubsection{An intervention}
Next, we consider introducing a random swap between the first and second positions in the submitted list, with a small probability that is nonetheless sufficient to invalidate the compliance test.
Given the monotonicity properties established below, a complete invalidation (i.e., for any $k$ and $e$) is achieved if and only if the swap probability $\epsilon$ is large enough such that the type I error for first-position listing—when no other programs exert pressure—exceeds the significance level $\alpha$, that is, $Q^{a,0,\epsilon}_{i\to1}\equiv\lim_{e\to0}Q^{a,e,\epsilon}_{i\to1}>\alpha$.
\newline

\noindent{\bf Remark 3.}
{\it $Q^{a,e,\epsilon}_{i\to k}\geq Q^{a,0,\epsilon}_{i\to1}%=a\epsilon
$ irrespective of $e$ and $k$.
Moreover, $Q^{a,0,\epsilon}_{i\to1}$ is increasing in $\epsilon$.}
\newline

\noindent{\bf Proof.}
For the former, since $(\tilde\pi^\epsilon_1(\bm{l}),\tilde\pi^\epsilon_2(\bm{l}))=(l_1,l_2)$ with probability $\epsilon$ and $(l_2,l_1)$ with probability $1-\epsilon$, $Q^{a,e,\epsilon}_{i\to1}=\epsilon Q^{a,e,0}_{i\to2}+(1-\epsilon)Q^{a,e,0}_{i\to1}$ and $Q^{a,e,\epsilon}_{i\to2}=\epsilon Q^{a,e,0}_{i\to1}+(1-\epsilon)Q^{a,e,0}_{i\to2}$.
Moreover, since $\{\tilde\pi^\epsilon_1(\bm{l}),\dots,\tilde\pi^\epsilon_{k-1}(\bm{l})\}=\{l_1,\dots,l_{k-1}\}$ for $k>2$, $Q^{a,e,\epsilon}_{i\to k}=Q^{a,e,0}_{i\to k}$ for $k>2$.
Then, since by (\ref{qae0}), $Q^{a,e,0}_{i\to k}\geq Q^{a,e,0}_{i\to2}\geq Q^{a,e,0}_{i\to1}$ for $k>2$, by $\epsilon\leq1/2$, $Q^{a,e,\epsilon}_{i\to k}\geq Q^{a,e,\epsilon}_{i\to1}$ for $k\geq2$.
Moreover, again by (\ref{qae0}), $Q^{a,e,\epsilon}_{i\to1}\geq Q^{a,0,\epsilon}_{i\to1}%=1-\epsilon(1-a)-(1-\epsilon)=a\epsilon
$.

For the latter, by (\ref{qae0}), $Q^{a,0,\epsilon}_{i\to1}=1-\epsilon(1-a)-(1-\epsilon)=a\epsilon$.\hfill$\Box$
\newline

At the same time, since the exchange itself causes a welfare loss, even if the negation is achieved, the associated rank loss by the permutation intervention,
\begin{equation}
PRL^{a,\epsilon}\equiv E\left[\tilde\mu^\epsilon\left((1,\dots,L),\tilde{A}^a\right)-\mu\left((1,\dots,L),\tilde{A}^a\right)\right],\nonumber
\end{equation}
may also be substantial.
\newline

\noindent{\bf Remark 4.} {\it $PRL^{a,\epsilon}$ is also increasing in $\epsilon$.}
\newline

\noindent{\bf Proof.}
%For the latter,
%\begin{equation}
Note that
$PRL^{a,\epsilon}=(2-1)\Pr\left(\{1,2\}\subset\tilde{A}^a\right)\Pr\left(\tilde\pi^\epsilon_1((1,\dots,L))=2\right)=a^2\epsilon$.\hfill$\Box$
%\nonumber
%\end{equation}
\newline

Moreover, while the loss to yield a given type I error is smaller with a smaller acceptance rate, a smaller rate also more restricts the possibility of the negation% (within the range of $\epsilon\in[0,1/2]$)
.
\newline

\noindent{\bf Remark 5.} {\it $Q^{a,0,\epsilon}_{i\to1}=\alpha$ for some $\epsilon$ if and only if $\alpha\leq a/2$, which is increasing in $a$, while $PRL^{a,\epsilon}$ with $\epsilon$ solving $Q^{a,0,\epsilon}_{i\to1}=\alpha$ is also increasing in $a$.}
\newline

\noindent{\bf Proof.}
For the former, recall $Q^{a,0,\epsilon}_{i\to1}=a\epsilon$, which also implies the latter by $PRL^{a,\epsilon}=a^2\epsilon$.
\hfill$\Box$
\newline

Thus, whether in the first place, a sufficiently large type error can be achieved, and if so, with so allowable associated loss also remains the empirical issue.

\section{Empirical evaluation}
For each market, we set $(P_1,P_2,L)$ from the reported data (see Footnote 2 and Figures 1 and 2), derive $(a,e)$ from (\ref{a}) and (\ref{e}) with them, and calculate key statistics as the following table.
Here $RL^a_{\rm rand}$ denotes the rank loss if the matched program is randomly chosen from those which would accept the doctor,
\begin{equation}
RL^a_{\rm rand}=\left(1-\Pr\left(\tilde{A}^a=\emptyset\right)\right)\left(\frac{L+1}2-E\left[\mu\left((1,\dots,L),\tilde{A}^a\right)\left|\tilde{A}^a\neq\emptyset\right.\right]\right),\nonumber
\end{equation}
and $PRL^a_\alpha$ denotes the associated permutation rank loss in order to attain the type I error $\alpha$, i.e., with $a\epsilon=\alpha$,
\begin{equation}
PRL^a_\alpha=PRL^{a,\alpha/a}.\nonumber
\end{equation}

Note that the average list length in Japan was 3.25 in 2019, and therefore we set $L=3$ in the following estimation. However, there is a much larger room for the choice of $L$ since it is very heterogeneous in the data. 
Moreover, the impact of the length of the list on each estimate is greater when the length of the list is shorter. So we also calculate the key statistics for the case of $L=4$ for the robustness check.
\begin{center}
\begin{tabular}{|c||c|c|c||c|c||c|c||c|c||c|}\hline
Market&$P_1$&$P_2$&$L$&$a$&$e$&$RL$&$RL_{\rm rand}$&$\underline{Q}$&$PRL_{0.15}$&$QL$\\\hline\hline
U.S.&$0.5$&$0.3$&10&$0.3232$&$0.1024$&$0.7740$&$2.5588$&$0.4829$&$0.0485$&$0.0704$\\\hline
Japan (L=3)&$0.7$&$0.5$&3&$0.5783$&$0.2707$&$0.2053$&$0.4754$&$0.6443$&$0.0867$&$0.0513$\\\hline\hline
Japan (L=4)&$0.7$&$0.5$&4&$0.5599$&$0.2141$&$0.3425$&$0.8373$&$0.6591$&$0.0840$&$0.0685$\\\hline

\end{tabular}
\end{center}

Thus, the rank loss reaches more than 30\% ($0.7740/2.5588\approx0.302$) of what the efficient assignment adds to the random assignment.
Moreover, we can achieve the type I error of 15\% ($<0.3232/2\approx0.162$)\footnote{That is, %to achieve the type I error of 15\%, 
we can choose $\varepsilon$ such that $\varepsilon\leq 1/2$ if $\alpha=0.15$ and $a=0.3232$, from the fact that $a\epsilon=\alpha$.}, with the associated loss less than one-half ($0.0867/0.2053\approx0.4223$)
%a quarter ($0.0840/0.3425\approx0.245$) 
of the improvement by the negation.\footnote{Usually, a significance level of 5\% or 10\% is employed, but here we use 15\%. This is to consider the effectiveness of the intervention methods we will examine later, even under conditions where an applicant who promised to write the pressured program first on their list is more likely to be judged a betrayer if they are not matched.}

%以下の説明はイントロの繰り返しになっている。
As we can see from the figures in the table, the intervention of list replacement can reduce the expected rank loss to 42\% (from 0.2053 to 0.0867) of its original value in Japan and to 6\% (from 0.7740 to 0.0485) in the U.S.
%Why does the intervention have 
A greater effect in the U.S. than in Japan may be understood as follows. First, the list is longer in the U.S., so the rank loss due to pressure is greater. And because the list is longer in the U.S., the relative damage of swapping the first and second places is smaller. Other values of $a$ and $e$ may differ, but this shows that the replacement intervention is more effective in matching markets with longer average lists.

For the rank loss evaluation, consider that the match value of each program for the doctor is independently drawn from an identical continuous distribution, and the preference order is determined according to these values. 
Then, as is easily calculated, one rank difference is translated into $1/(L+1)$ quantile difference, that is, the expected quantile loss is
\begin{equation}
QL^{a,e}=\frac{RL^{a,e}}{L+1},\nonumber
\end{equation}
which for each market is also reported in the table,
where we find that once evaluated in quantile, the estimated losses are similar.
Thus, on average, approximately 5 to 7 percent point quantile is lost in both markets.

\begin{comment}
We also plot how the estimated quantile loss would be if $P_2$ took different values, as follows.
Somewhat interestingly, the actual $P_2$ attains almost highest value in both markets.
\begin{center}
\includegraphics[width=7.5cm]{QLbyP2_US.png}
\includegraphics[width=7.5cm]{QLbyP2_JP.png}
\\
$QL$ (vertical) by $P_2$ (horizontal) for the U.S. (left) and Japan (right)
\end{center}
\end{comment}

Finally, it may also be interesting that the estimated pressure probability is much smaller ($0.1024\ll
%0.2141
0.2707$) in the U.S. market, where the matching authority explicitly prohibits such conduct.

\section{Discussion}
\subsection{Doctors' dissatisfaction with the intervention}
If an intervention is implemented that swaps the first and second-ranked hospitals in a doctor's rank-order list, and a doctor ultimately matches with the second-ranked hospital on their list, they may suspect that the intervention affected the matching outcome. In such cases, the doctor might contact the hospital they originally ranked first to inquire whether they were ranked highly enough on the hospital's list to have matched under normal circumstances. If the doctor finds that they would have matched with their first-choice hospital but instead matched with their second-choice hospital due to the intervention, they may express dissatisfaction with the outcome.

However, the potential loss from the intervention is limited: the doctor matches with their second rather than their first choice. The intervention does not cause doctors to match with hospitals ranked lower than second place. Furthermore, as will be discussed in Section 3, it is important to note that the current situation is sufficiently suboptimal that, on average, such an intervention would be beneficial. This assessment, however, assumes that rank losses are uniform—treating the gap between first and second choices as equal to that between second and third. In reality, some doctors may perceive the difference between their first and second choices as significantly greater than that between their second and third choices.

\subsection{The possibility of strategic manipulation}
We have examined how randomly permuting a doctor's submitted list can mitigate rank loss.
This raises the question: can doctors strategically alter their behavior in response to such an intervention?
The answer is no, provided that the permutation probability $\varepsilon$ is no more than one-half and that the probability that a program will accept a specific doctor is exogenously given by $a$, as assumed in this paper.\footnote{See also footnote 15 for the discussion of strategic manipulations.}

However, when applying the permutation intervention to real-world residency matching, strategic manipulation becomes a more pressing concern.
To illustrate, consider the situation without the constant acceptance rate assumption.
Suppose that each program's acceptance probabilities differ, and that a doctor is aware of these values. In this case, even if the clearinghouse swaps the first and second positions on a doctor's submitted list with a certain probability, the doctor can take actions that effectively nullify the permutation intervention.
For example, if a doctor identifies a program with a very low acceptance rate,  they can place that program in the top position as a decoy, thereby preserving their true preferences.
This manipulation ensures that the doctor's true first and second choices remain unchanged.

This example implies that strategic manipulation is possible in reality and that doctors do not always submit lists based on their true preferences, even after clearinghouse intervention.

An alternative intervention method proceeds as follows: First, a temporary match is computed using the DAA. Next, one of the applicants who was matched with a hospital is selected with an exogenous probability $\varepsilon (>0)$, and the matching algorithm is run again after removing the hospital that matched that applicant from the list of applicant submissions.

We can show that this mechanism works in cases where only one doctor's matched program is removed in the temporary match.\footnote{See appendix available on request.}
Moreover, further study is needed on the potential for strategic manipulation under this mechanism.

\subsection{Stability}
This paper examined the effect of independently swapping the first and second positions in each applicant's submitted list with probability $\varepsilon$.
This intervention creates the possibility of a blocking pair; therefore, the modified matching algorithm is no longer stable.

Stability is a key concept in matching theory. 
However, one might argue that stability is not as crucial in the context of residency matching.
This is because the clearinghouse can enforce the outcome, preventing participants from acting on blocking pairs.
Moreover, as the training period is typically short (e.g., two years), doctors may accept suboptimal matches.

In addition, even in the current matching system where pressure is present, a stable matching has not been achieved.
Our analysis shows that, when comparing two types of instability—one caused by external pressure in the current system, and the other resulting from intervention side effects—the latter leads to smaller welfare losses.

For the above reasons, stability may appear less important in residency matching. However, as discussed later, matching theory is applied in a wide range of areas, including assignments within firms.
Therefore, while stability may be less critical in residency matching, it remains essential in broader applications of matching theory and merits further investigation.

\subsection{Welfare consequences for hospitals}
We have focused on a representative doctor and evaluated interventions designed to reduce rank loss caused by pressure from hospitals to be ranked first. In doing so, we assessed welfare based on the doctor's rank loss.

To fully evaluate the benefits of clearinghouse interventions, it is also essential to consider hospital welfare.
If there is no variation in the value among acceptable doctors for a hospital, then welfare can be evaluated solely by focusing on the doctor's rank loss.
Moreover, suppose there exists a value correlation such that a program that a particular resident highly values will also highly value that doctor. 
In the extreme case of perfect correlation, the doctor's rank loss alone suffices to evaluate welfare.

However, the value of accepting a doctor may vary significantly across hospitals. For example, a hospital in a doctor-shortage area may have a greater need to accept a resident than a hospital in an urban area.
Consequently, hospitals with a greater need to accept a resident may exert pressure on applicants.
In such cases, pressure from hospitals on doctors may, under certain conditions, contribute positively to social welfare.
Thus, the social welfare impact of a program that exerts pressure on doctors depends on the relative magnitude of the rank loss experienced by doctors and the gains for hospitals.

\section{Conclusion}
This paper investigated an anomaly in conditional match rates and its underlying causes.
First, we showed that this anomaly can be explained by programs pressuring applicants to rank them first in order to ensure acceptance.
Then, using a simple reduced-form model of acceptance and pressure, we demonstrated that data from the U.S. and Japan imply significant pressure and associated welfare losses.
The probability that a program exerts pressure on an applicant is more than twice as high in Japan (0.2707) as in the U.S. (0.1024), likely due to weaker (or absent) direct regulations against pressure in Japan.

Finally, we proposed a potential solution: randomly permuting applicants' submitted lists before running the matching algorithm. This intervention may effectively neutralize pressure.

The method we propose for estimating pressure probabilities can be applied to matching markets beyond residency assignments.
In recent years, matching algorithms have been used in various environments.
For example, Google employs algorithms to determine worker assignments.\footnote{See for detail ``Google's algorithm-powered internal job marketplace'' (\url{https://rework.withgoogle.com/blog/googles-algorithm-powered-internal-job-marketplace/}).} 
However, in internal labor markets, participants may still attempt to influence outcomes by pressuring one another.
This highlights the need for further research into the inefficiencies caused by pressure and the development of effective countermeasures.

%\newpage

\end{document}